\documentclass[12pt, column,
showpacs,preprintnumbers,amsmath,amssymb,nodata]{revtex4}

\usepackage{epsfig}
\usepackage{graphics}

\newcommand{\be}{\begin{equation}}
\newcommand{\ee}{\end{equation}}
\newcommand{\bn}{\begin{eqnarray}}
\newcommand{\en}{\end{eqnarray}}

\usepackage[english]{babel}
\begin{document}

\title{Relaxation and phase space singularities  in time series of human
 magnetoencephalograms  as indicator  of  photosensitive epilepsy}
\author{\firstname{R.~M.}~\surname{Yulmetyev}$^{1, 2}$}
\email{rmy@theory.kazan-spu.ru}
\author{\firstname{P.}~\surname{H\"anggi}$^{3}$}
\author{\firstname{D.~G.}~\surname{Yulmetyeva}$^{1, 2}$}
\author{\firstname{S.}~\surname{Shimojo}$^{4}$}
\author{\firstname{E.~V.}~\surname{Khusaenova}$^{1, 2}$}
\author{\firstname{K.}~\surname{Watanabe}$^{5, 6}$}
\author{\firstname{J.}~\surname{Bhattacharya}$^{7, 8}$}

\affiliation{$^{1}$Department of Physics, Kazan State University,
Kremlevskaya Street, 18 Kazan, 420008 Russia\\$^{2}$Department of
Physics, Kazan State Pedagogical University,  Mezhlauk Street, 1
Kazan, 420021 Russia} \affiliation{$^{3}$Department of
Physics,University of Augsburg, Universit\"atsstrasse 1, D-86135
Augsburg, Germany} \affiliation{$^{4}$Division of Biology,
CalTech, Pasadena, CA 91125 USA}
 \affiliation{$^{5}$ Research Group for Decision Making,
 Research Center for Advanced Science and
Technology,  University of Tokyo, Japan}
 \affiliation{$^{6} $ERATO Shimojo Implicit Brain Function Project,
 Japan Science and Technology Agency, Japan}
 \affiliation{$^{7} $ Comission for Scientific
Visualisation, Austrian Academy of Sciences, Tech Gate, Vienna A -
1220, Austria} \affiliation{$^{8}$ Department of Psychology,
Goldsmiths College, University of London, New Cross, London SE14
6NW, UK}

\begin {abstract}
To analyze  the crucial role of fluctuation and relaxation effects
for the function of the human brain  we  studied some statistical
quantifiers that support the information characteristics of
neuromagnetic brain responses (magnetoencephalogram, MEG). The
signals to a flickering stimulus of different color combinations
have been obtained from a group of control subjects which is then
contrasted with those of a patient suffering photosensitive
epilepsy (PSE). We found that the existence of the specific
stratification of the phase clouds and the concomitant relaxation
singularities of the corresponding nonequilibrium dynamics  of the
chaotic behavior of the signals in separate areas in a patient
provide likely indicators for the zones which are responsible for
the appearance of PSE.

\end {abstract}
\pacs{05. 45. Tp; 87. 19. La; 89. 75. -k}
 \maketitle

\section{Introduction}
The study of dynamical time series  is gaining ever increasing
interest and is
 applied and used  in diversified
  fields of natural
sciences, technology, physiology, medicine and economics \cite{Jade,
Barb, Rag, Mar, Lau, Hir, Lau2,kantzschreiber}, to name only a few.
The majority of natural systems can be considered dynamical systems,
whose evolution can be studied by time series related to relevant
variables on a suitable time scale. These series are often
characterized by a pronounced time and spatial synchronization or
coherence, chaotic or robust behavior.

When analyzing time series data with linear methods, one can follow
certain standard procedures. Moreover, the behavior may be
 described by a relatively small set of parameters. For a
nonlinear time series analysis, this is not necessarily the case.
While standardized algorithms exist for the analysis of the time
series data with nonlinear methods, the application of these
algorithms requires considerable knowledge and skills on the part of
the user.

In a nonlinear time series analysis one starts out with a
reconstruction of the state spaces from the  observed data
\cite{kantzschreiber,Tak, Sau, Stark, Sau2}. Although   the
embedding theorems \cite{Schrei} provide an important means of
understanding the reconstruction procedure, likely none of them is
formally applicable in practice. The reason is that they all deal
with infinite, noise free trajectories of a dynamical system. It is
not obvious that the theorems should be "approximately valid"  if
only the requirements are "approximately fulfilled", for example, if
the data sequence is long, although  finite and  not completely
noise-free.

A possible way to study  the manifestation of physical properties of
random processes (and the Markov random processes (MRP) in
particular) in time series originates from the theory of
nonequilibrium statistical physics. The history of the fundamental
role of stochastic processes in physics dates back a century to the
Markov representations  \cite{Mark} of random telegraphic signals.
Such processes still find application in models of contemporary
complex phenomena. A few typical examples of complex physical
phenomena modeled by the Markov stochastic processes are: kinetic
and relaxation processes in gases \cite{gases} and plasma
\cite{plasma}, condensed matter physics (liquids \cite{liq}, solids
\cite{solids}, and superconductivity \cite{super}), astrophysics
\cite{astro}, nuclear physics \cite{nucl}, for certain quantum
relaxation dynamics \cite{quant} and in classical \cite{class}
physics. At present, we can make use of a variety of  statistical
methods for the analysis of the Markov and the non-Markov
statistical effects in diverse physical systems. Typical examples of
such schemes are the Zwanzig-Mori's kinetic equations \cite{Zwanz},
generalized master equations and corresponding statistical
quantifiers \cite{hanggi}, the Lee's recurrence relation method
\cite{Lee}, the generalized Langevin equation (GLE) \cite{GLE}, etc.

In this paper we shall study the crucial role of relaxation and
kinetic singularities in brain function  of healthy physiological
and pathological systems for the case of photosensitive epilepsy
(PSE). In particular, we seek marked differences in large space and
times scales in the corresponding stochastic dynamics of discrete
time series that could in principle  characterize  pathological (or
catastrophic) violation of salutary dynamic states of the human
brain. As a main result,  we show here that the appearance of
distinct differences in the  relaxation time scales and
extraordinary stratification of the phase clouds in the stochastic
evolution of neuromagnetic responses of the human brain as recorded
by MEG may yield  evidence of  pronounced zones responsible for the
appearance of PSE.\\

\section{Stratification in the phase space
 and  stochastic processes in complex systems}

 The phase space  plays a crucial role in determining
 the singularities of stochastic dynamics of the underlying  system.
  A set of the dynamical orthogonal
   statistical variables
 describing  the dynamical state of the complex system is a feature important
 in a proper construction of the phase space and analysis of the
 underlying dynamics. Let us consider  an k-dimensional   vector of
 the initial state
 $ {\bf A}_k^0 = À (x_1, x_2 , x_3 , ....  x_ k $) and an k-dimensional  dynamic vector of
 the final state
 $ {\bf A}_{k + m}^m = À (x_{m+1}, x_{m+2 }, x_{m+3 }, ....  x_{ m+k} $) ,  where
$ k + m = N ,  k,   m  =  N, N  - 1, N  - 2,... ,N/2 - 2, N/2 - 1,
N/2 $ and N denotes the sample length. From the discrete equation of
motion

\be \nonumber
\frac{\Delta x_i}{\Delta
t}=\frac{x_{i+1}-x_i}{t_{i+1}-t_i}=\frac{1}{\tau}\left\{\Delta-1\right\}x_i,
\ee
\be 
t_{i+1}-t_i=\tau,
\ee
 we obtain the equation of motion of the dynamical vectors
of state $ {\bf A}_{ m+k}^m$ as:

\be \nonumber
\frac{\Delta { \bf A}_j^m}{\Delta t}=i\widehat{\emph{L}}A_j^m, \ee
\be 
i\widehat{\emph{L}}= \frac{1}{\tau}{(\Delta-1)} \;, \ee

where the shift operator $ \Delta $ acts as $ \Delta x_j= x_{j+1}$
and j = m+k.

By applying successively  the quasioperator $\widehat{\emph{L}}$
to the dynamic variables $A_j^m(t)$, t=m$\tau,$  where $\tau$ is a
discrete time step, $j = m+k $,  we obtain the  set of dynamic
functions ${\bf B}_n(0)= \widehat{\emph{L}}^n {\bf A}_k^0(0)$,
$n>1$.
 By using the variables $B_n(0)$ one can find a
formal solution of the  evolution of these equations in the form

\be 
{\bf A}_{m+k}^m(m\tau)=\left\{1+i\tau\widehat{\emph{L}}\right\}^m
{\bf A}_k^0(0) =\sum_{j=0}^{m}\frac{m!(i\tau)^{m-j}}{j!(m-j)!}{\bf
B}_{m-j}^0.
\ee

However, the use of this structure is generally  not the most
convenient one. An advantageous approach consists in the use of  the
{\it orthogonal}  vectors $W_n$, as detailed below, by use of the
Gram-Schmidt orthogonalization procedure \cite{Reed} in place of the
set of variables $B_n(0)$.  Thus, we work with this new  set of
dynamical orthogonal vector variables, where the average $<...>$
should be read in terms of scalar products
 and $\delta_{n,m}$ is the Kronecker's symbol. In doing so, we find
 the the recurrence formula in which the "senior" variables at order
 "n" $W_n=W_n(t)$ are related through with the "junior" variables of
 order $m <n$; i. e. \cite{DFA,Yulm}

\be \nonumber
{\bf W} _0 = {\bf A}_k^0 (0), ~~{\bf W}_1 = \{ i \hat L-\lambda_1
\}{\bf W}_0,\ldots ~~ \\\nonumber
\ee
\be \nonumber
{\langle  {\bf W}_n  {\bf W}_m \rangle} = \delta_{n,m}{\langle|\bf
W_n|^2\rangle},~~
\\\nonumber
\ee
 \be 
 {\bf W}_n = \{ i\hat L-\lambda _ n\} {\bf W}_{n-1}+
\Lambda_{n-1} {\bf W}_ {n-2}+..., ~~ n > 1 . \label {f3} \ee Here,
we have used the kinetic  parameters given by  the Liouville's
quasioperator L as follows \cite{Yulm}:

\be \nonumber
\lambda_n=\frac{\langle {\bf W_n} \widehat{\emph{L}} {\bf W_n}
 \rangle}{\langle|{\bf W_n}|^2\rangle} ,
\ee
\be 
\Lambda_n=\frac{\langle|{\bf W_n}|^2\rangle}{\langle|{\bf
W_{n-1}}|^2\rangle} , \ee where $ \Lambda_n =  \Omega_{n-1}^2 $,
with  the parameter  $\Omega_n $ denoting a general relaxation
frequency. The set of frequencies $\lambda_n$ describes the spectrum
 of the Liouville's operator $\widehat{\emph L}$.

Besides these relaxation time  measures, we in addition shall
consider information measures that are based on time correlation
functions which assume the role of memory functions $M_{i}(t), i =
0, 1, 2 , ...$, see in Refs. \cite{Yulm}; i. e.,

\be \nonumber
M_i (t) = \frac{\langle {\bf W_i(0)}  {\bf W_i (t)}
 \rangle}{\langle|{\bf W_i(0)}|^2\rangle}   .
\ee Note that the quantity with $i = 0$ corresponds to the temporal
autocorrelation function of the vector ${\bf A}_k^0$.

\section{Analysis of time series for the experimental data of PSE  }

Now we can proceed directly to the analysis of the experimental
data: the MEG signals recorded in a group of nine healthy human
subjects and  a patient with (PSE) \cite{MEG}. PSE is a common
type of stimulus-induced epilepsy, defined as recurrent
convulsions precipitated by visual stimuli, particularly
flickering light. The diagnosis of PSE involves the detection of
paroxysmal spikes on an EEG in response to the intermittent light
stimulation. To elucidate the color-dependency of PS in normal
subjects,  brain activities subjected to uniform chromatic
flickers with whole-scalp MEG have been measured in Ref.
\cite{MEG}).

The same subjects and the data set were part of an earlier study
in Ref. \cite{MEG}; however, we shall mention the relevant details
for the sake of completeness. Nine healthy adults (two females,
seven males; with age ranging from 22-27years) voluntarily
participated. Two additional age-matched child control subjects,
and one more photosensitive patient (age 14yr) under medication
(sodium valporate), were also studied. All subjects were
right-handed and were explicitly informed that flicker stimulation
might lead to epileptic seizures. They gave their written informed
consent before recording. The subjects were instructed to
passively observe visual stimuli with minimal eye movement. During
the testing session for the photosensitive patient, pediatric
neurologists were present to monitor their health condition as a
precautionary measure.

\begin{figure}[h!]
\leavevmode \centering
\includegraphics[height=10cm,width=12.5cm,angle=0]{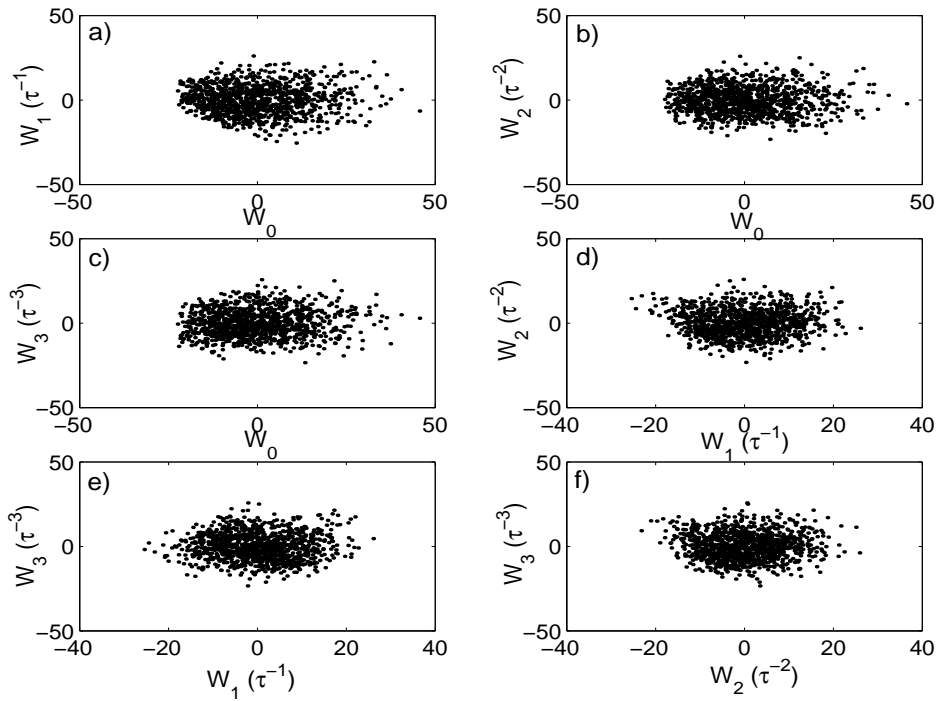}
\caption{The single  phase planes of the phase portrait of the
patient with PSE for MEG signals from an R/B combination of the
light stimulus : a) plane $(W_0 (t), W_1(t))$; b)  plane $(W_0(t),
W_2(t)$); c)  plane $(W_0(t), W_3(t)$); d)  plane $(W_1(t),
W_2(t)$); e) plane $(W_1(t), W_3(t)$); f)  plane $(W_2(t),
W_3(t))$.}
\end{figure}



 The subjects were screened
for photosensitivity and personal or family history of epilepsy.
The experimental procedures followed the Declaration of Helsinki
and were approved by the National Children's Hospital in Japan.
The stimuli were generated by the two video projectors and
delivered to the viewing window in the shield room through an
optical fiber bundle. Each projector continuously produced a
single color stimulus. Liquid crystal shutters were located
between the optical device and the projectors. By alternative
opening one of the shutters for 50 ms, 10 Hz (square-wave)
chromatic flicker was produced at the viewing distance of 30 cm.
Three color combination were used : red-green (R/G), blue-green
(B/G),  and red-blue (R/B). CIE coordinates were x=0. 496, y=0.
396 for red; x=0. 308, y=0. 522 for green; and x=0. 153,  y= 0.
122 for blue. All color stimuli had the luminance of $1. 6$
cd/m$^{2}$ in otherwise total darkness. In a single trial, the
stimulus was presented for 2s and followed by an inter-trial
interval of 3s, during which no visual stimulus was displayed. In
a single session, the color combination was fixed.

\begin{figure}[h!]
\leavevmode \centering
\includegraphics[height=9cm,width=14.5cm,angle=0]{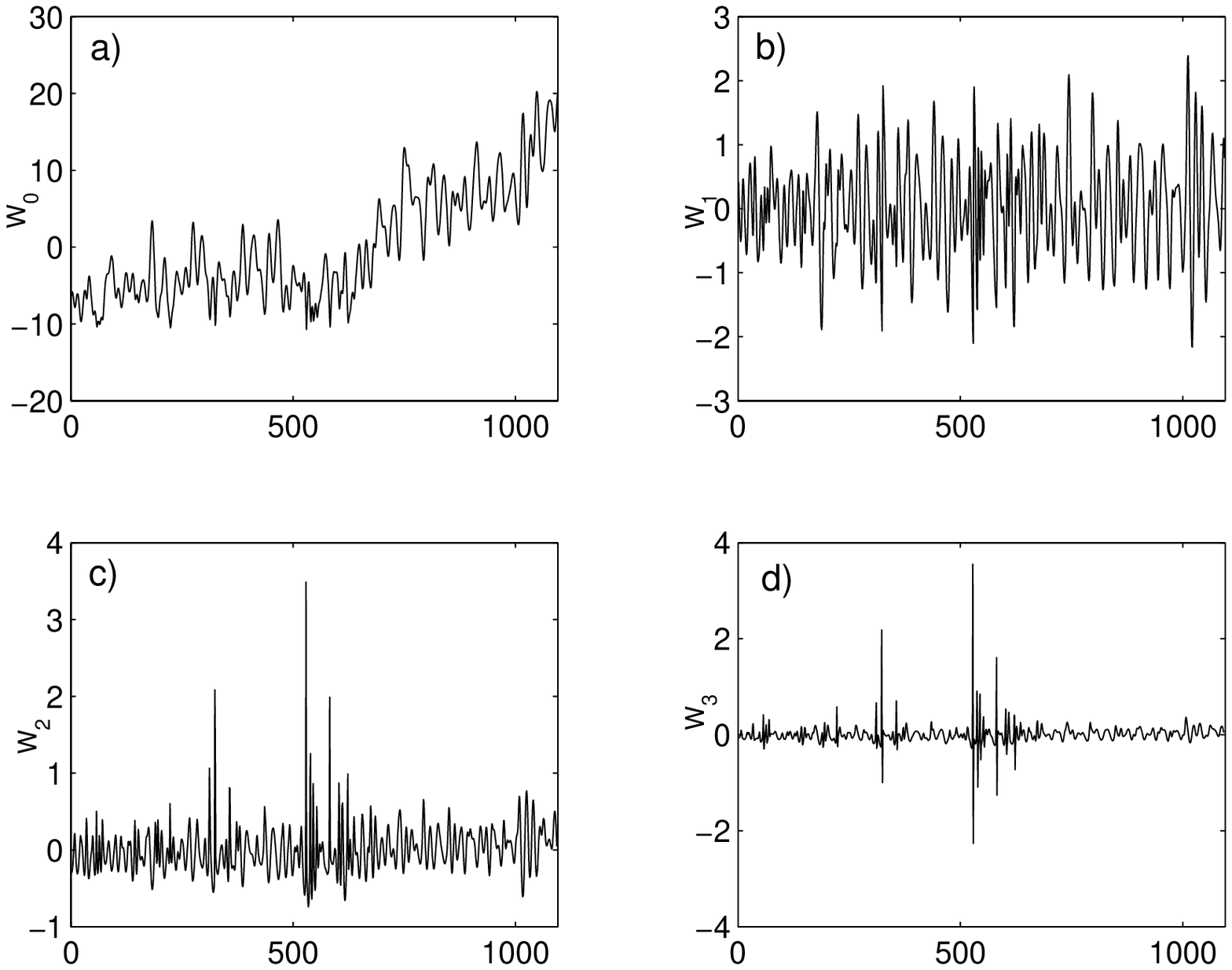}
\caption{The time dependence of the first four orthogonal variables
for MEG signals in  healthy in No. 3    ( sensor m = 13, for an R/ B
combination of the light stimulus: for healthy subject No. 4 : a)
$W_ 0(t)$; b) $W_ 1(t)$; c) $W_ 2(t)$;d) $W_ 3(t)$ for an  R/ B
combination of the light stimulus.}
\end{figure}

Neuromagnetic responses were measured with a 122-channel
whole-scalp neoromagnetometer (Neuromag - 122; Neuromag Ltd.
Finland). The neoromag-122 has 61 sensor locations, each
containing two originally oriented planner gradiometers coupled
with dc-SCUID (superconducting quantum interference device)
sensors. The two sensors of each location measure two orthogonal
tangential derivatives of the brain magnetic field component
perpendicular to the surface of the sensor array. The planner
gradiometers measure the strongest magnetic signals directly above
local cortical currents. From 200 ms prior responses were
analog-filtered (bandpass frequency 0.03 - 100 Hz) and digitized
at 0.5 kHz. Eye movements and blinks were monitored by measuring
an electro-oculogram.

\begin{figure}[h!]
\leavevmode \centering
\includegraphics[height=10cm,width=14.5cm,angle=0]{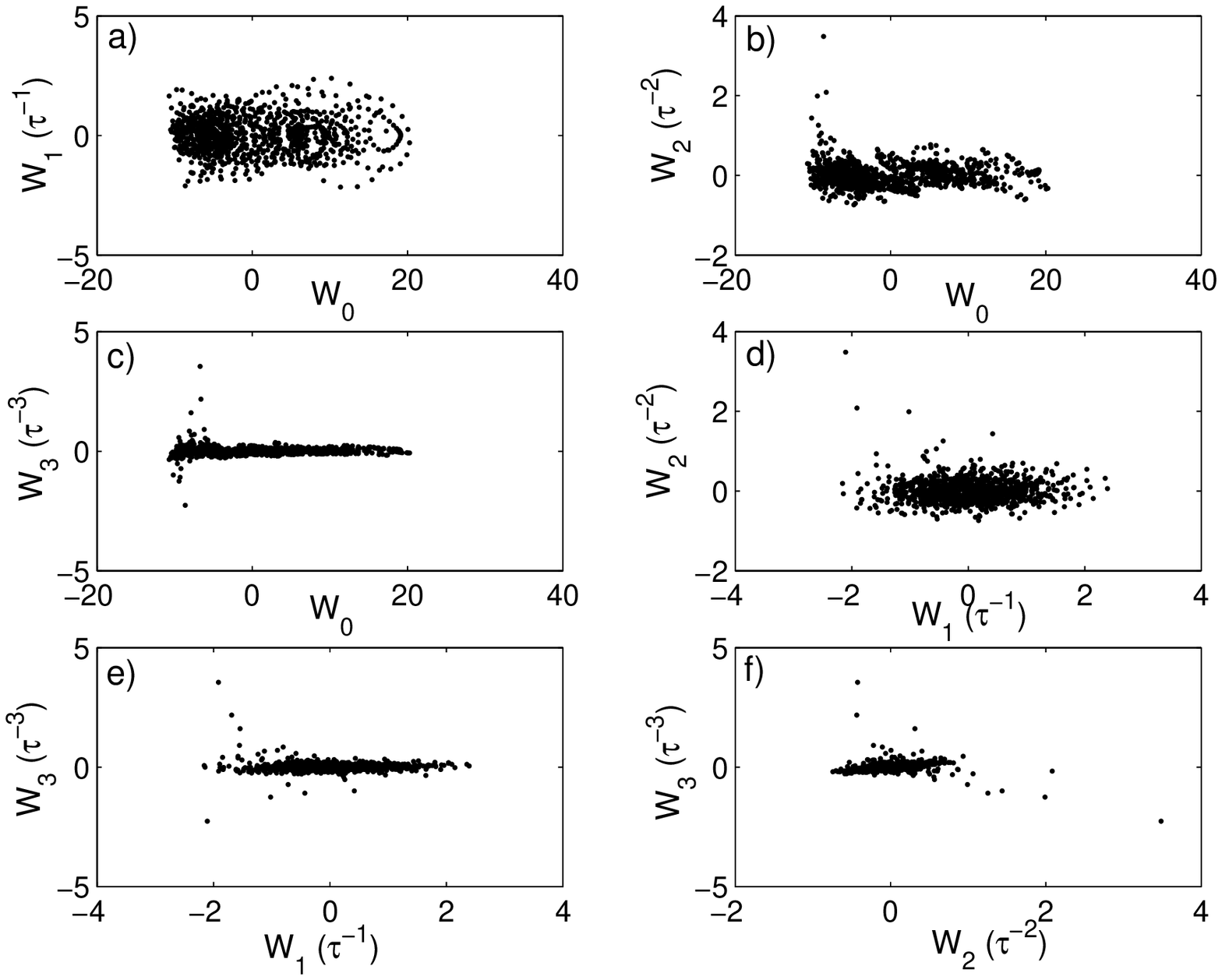}
\caption{The phase portraits  for MEG signals in healthy person No.
3 (sensor number m = 13) for an R/ B combination of the light
stimulus. The phase space has been created by the phase points
$\Gamma_{ i, j}(t) = (W_ i(t), W_ j(t)), 1 \leq i \neq j \leq3$.}
\end{figure}

The trials with MEG amplitudes $> 3000$ fT/cm and/or
electro-oculogram amplitudes $> 150 \mu$ V were automatically
rejected from averaging. The trials were repeated until $ >80$
responses were averaged for each color-combination. The averaged
MEG signals were digitally lowpass-filtered at 40 Hz, and then the
DC offset along the baseline $(-100$ to $ 0 $ ms) was removed. At
each sensor location, the magnetic waveform amplitude was
calculated as the vector sum of the orthogonal components. The
peak amplitude were normalized within each subject with respect to
the subject's maximum amplitude. The latency range from $ -100$ to
$-1100$ ms was divided into 100 ms bins. Then, the peak amplitudes
were calculated by averaging all peak amplitudes within each bin.
It would be important to mention that no clinical photosensitive
seizures were induced during the experiment. This also confirms
the better detection power of this analysis than normal seizure
detection procedure.

\begin{figure}[h!]
\leavevmode \centering
\includegraphics[height=10.5cm,width=14cm,angle=0]{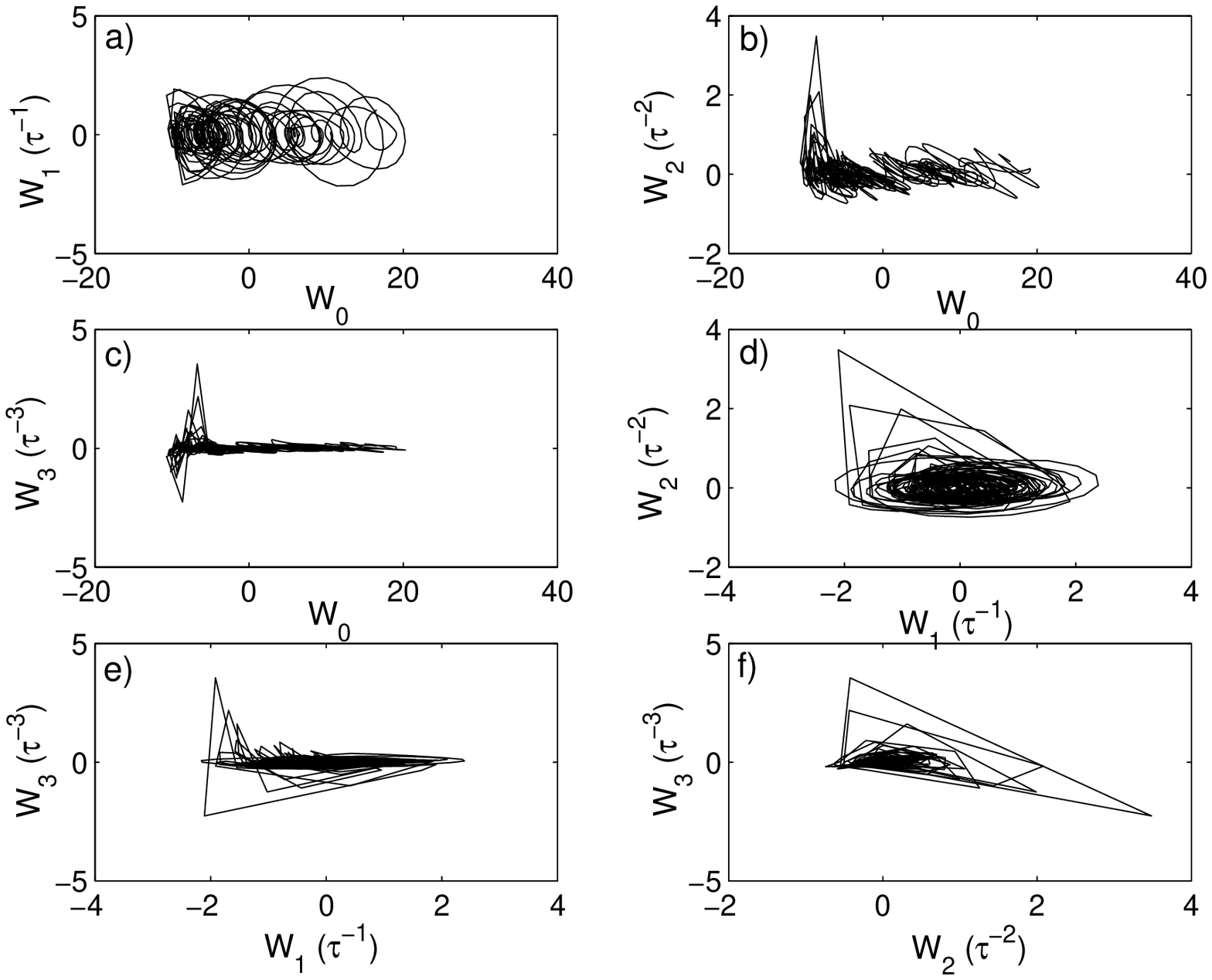}
\caption{The phase trajectories of the phase points $\Gamma_{i,j}(t)
= (W_i(t), W_j(t))$,
 healthy person No. 3  (sensor number m = 13) for an R/ B combination of the
light stimulus.}
\end{figure}

\section{Results and discussion }

Here we present the information-theoretic analysis for the presence
of PSE, based on the time behavior of the dynamical variables and
the stratified structure of phase spaces. Some results of our
quantities as derived from the theory presented here, are depicted
in Figs. 1-7. Our results for nine healthy subjects and for the
patient with PSE in comparison are shown in Figs. 8 - 10 . Among
them are: the time traces of the MEG signals $(W_0)$, and for the
three junior dynamical orthogonal variables $(W_i)$, i = 1, 2 and 3;
2), the phase space created by the points with coordinates $(W_i)$,
i = 0, 1, 2 and 3; 3)  the phase space, filled by the trajectories
$(W_i(t))$, i = 0, 1, 2 and 3; 4) the time dependence of the first
four quantifiers: the  time correlation functions (TCF) $M_0(t)$ and
the first three junior memory functions (MF) $M_i(t)$, for i = 1, 2
and 3. The results of the experiment for a red-blue (R/B) and a
red-green (R/G) combination of color signals are used in all of the
figures.

\begin{figure}[h!]
\leavevmode \centering
\includegraphics[height=8cm,width=14cm, angle=0]{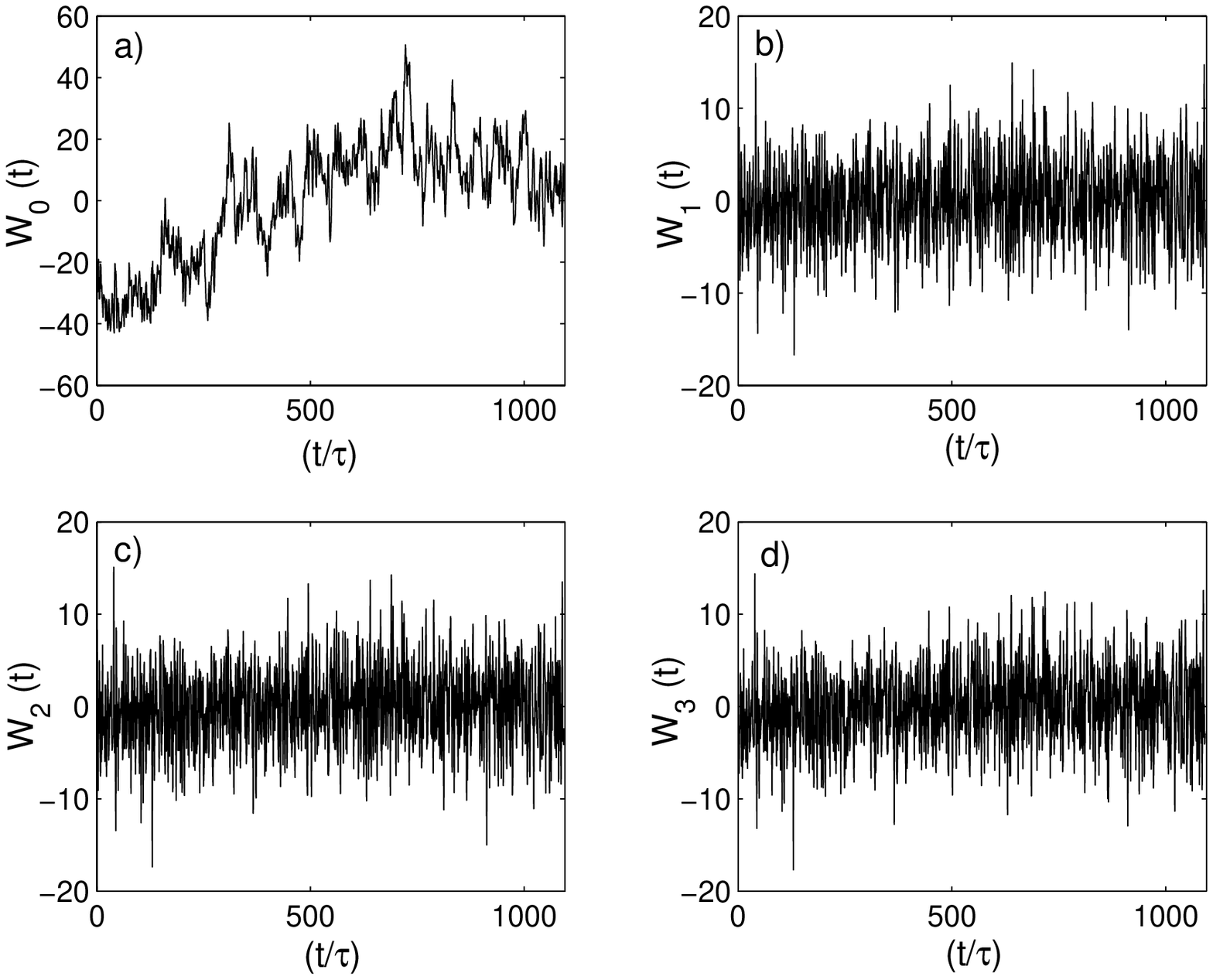}
\caption{The time dependence of the first  four orthogonal variables
$W_ i(t),  i=1, 2,...  3$ for the MEG signals for the patient with
PSE, sensor m = 13, an R/ B combination of the light stimulus.}
\end{figure}

  As an example, the  analogous
   results for the patient with PSE(sensor No. 10) are
   presented in Fig. (1). The obtained results possess
   the clearly visible inconsistent character. The comparison clearly shows
   a characteristic behavior of the dynamic
   variables $(W_i(t))$ (i = 1, 2 and 3).
   The time dependence of the variables $(W_1(t))$  presents  the time behavior of  the orthogonal velocity of the
   signal recording
   in  discrete form. The next-order dynamic variable $(W_2(t))$ describes the
   orthogonal acceleration, the variable $(W_3(t))$  depicts the
   longitudinal orthogonal energy current, etc. The signals
   $(W_i(t))$ in the patient with PSE can be characterized as regular noise.
    The phase clouds  formed by manifold
    phase points drastically differ
   from the healthy ones in comparison with
   the patient with PSE, see, Fig. 1.

\begin{figure}[h!]
\leavevmode \centering
\includegraphics[height=10cm,width=14cm,angle=0]{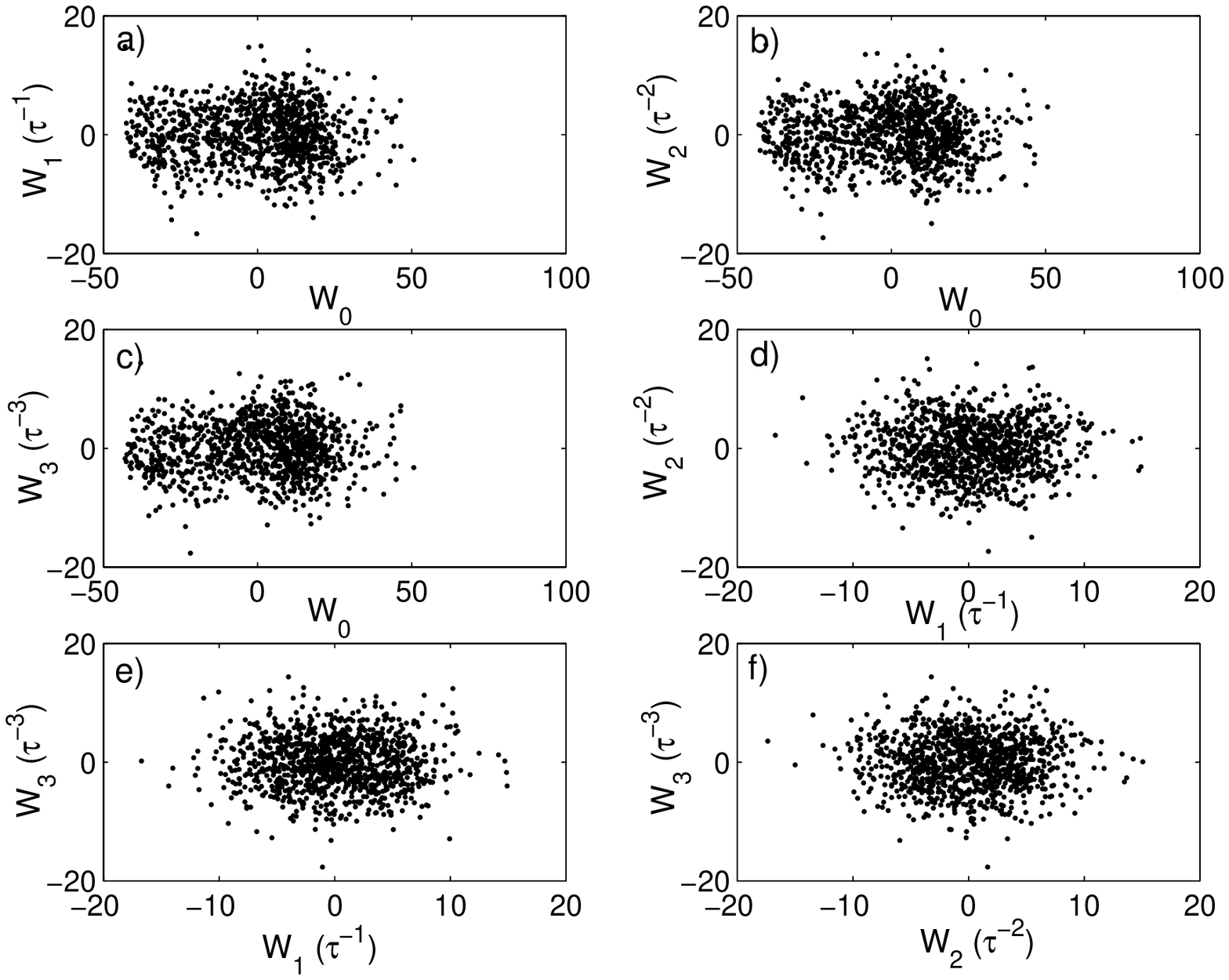}
\caption{The phase portraits created by the phase points
$\Gamma_{i,j}(t) = (W_i(t),W_j(t))$ from MEG signals for patient
with PSE, m = 13, an R/ B combination of  the light stimulus.}
\end{figure}


   The stratification of the
   phase clouds and the existence of the stable pseudoorbits are
   more visible for the healthy . In the patient with PSE (see,Fig.
   (1)) the phase stratification  disappears. The phase clouds
   can be  characterized by symmetrical nuclei, they have spatial
   homogeneity. The phase trajectories for the healthy
   form broken lines.
\begin{figure}[h!]
\leavevmode \centering
\includegraphics[height=9cm,width=14cm, angle=0]{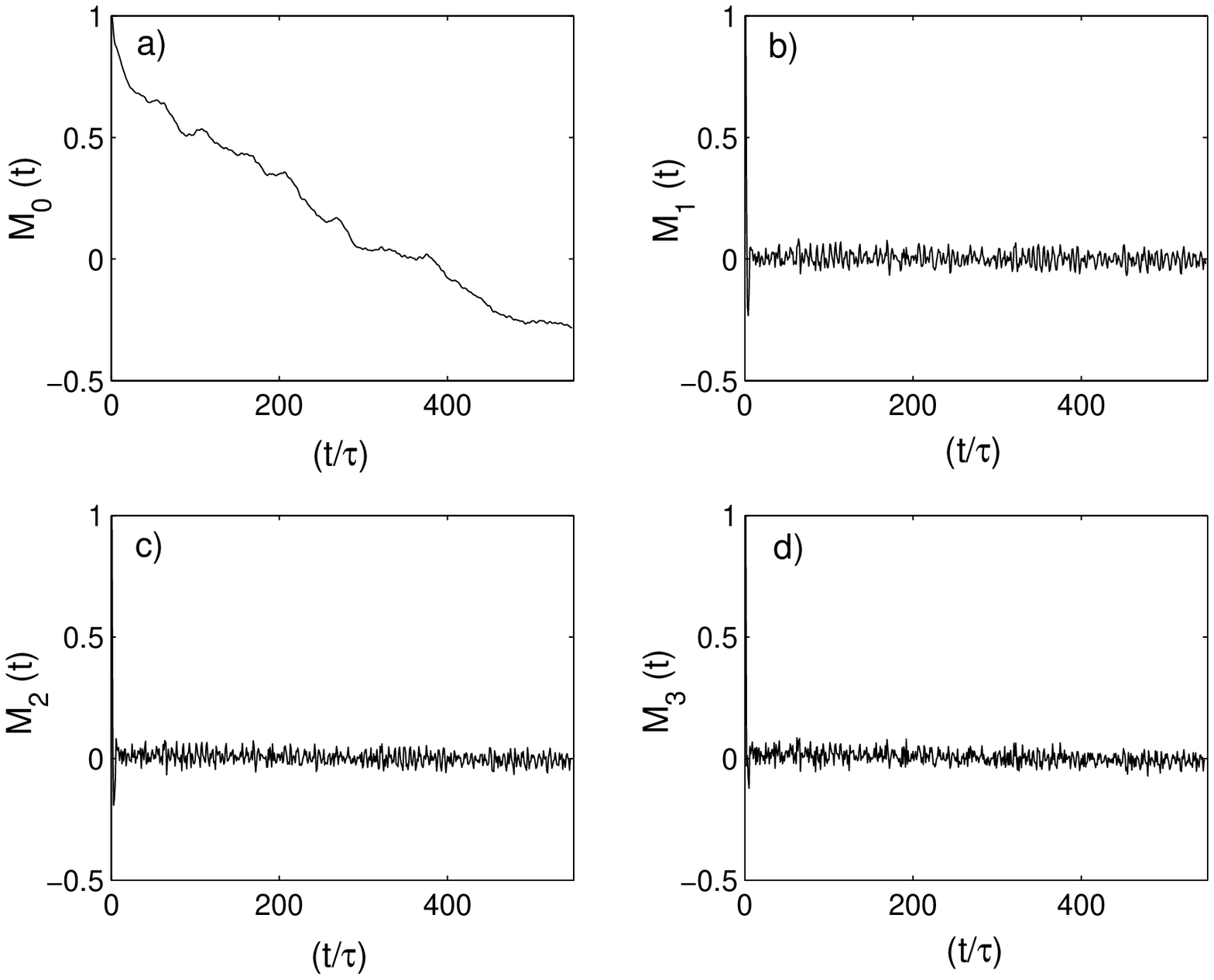}
\caption{The time dependence  of the first  four MF's  $M_ i(t), i=
0, 1,...  3$  for the MEG signals in the patient with PSE ( sensor
number m = 13) an R/ B combination of the light stimulus. Large
scale fluctuation of all functions become obvious in comparison with
a case for the healthy.}
\end{figure}

  For the patient with PSE the pictures of the phase
trajectories contrast sharply with the  case of the healthy. The
phase trajectories are packed tightly within the restricted areas
of the phase space. The drastic  difference in the typical scales
of the dynamic variables  $(W_i(t))$ and in the size of the phase
space for the healthy  and for the patient with PSE (Fig. 1) are
striking. This difference ranges from 3 times (for $(W_0(t))$ ) to
10 times(for $(W_i(t)), i = 1, 2,$ and  3) and from  10 times (for
the phase plane ${(W_0(t), W_1(t)})$ to 80 times in the
corresponding clouds of  phase planes ${(W_0(t), W_i(t)})$, i = 2,
3 and ${(W_i(t), W_j(t)})$ with $1\leq i, j\leq 3$.

\begin{figure}[h!]
\leavevmode \centering
\includegraphics[height=10.5cm, width=14cm,angle=0]{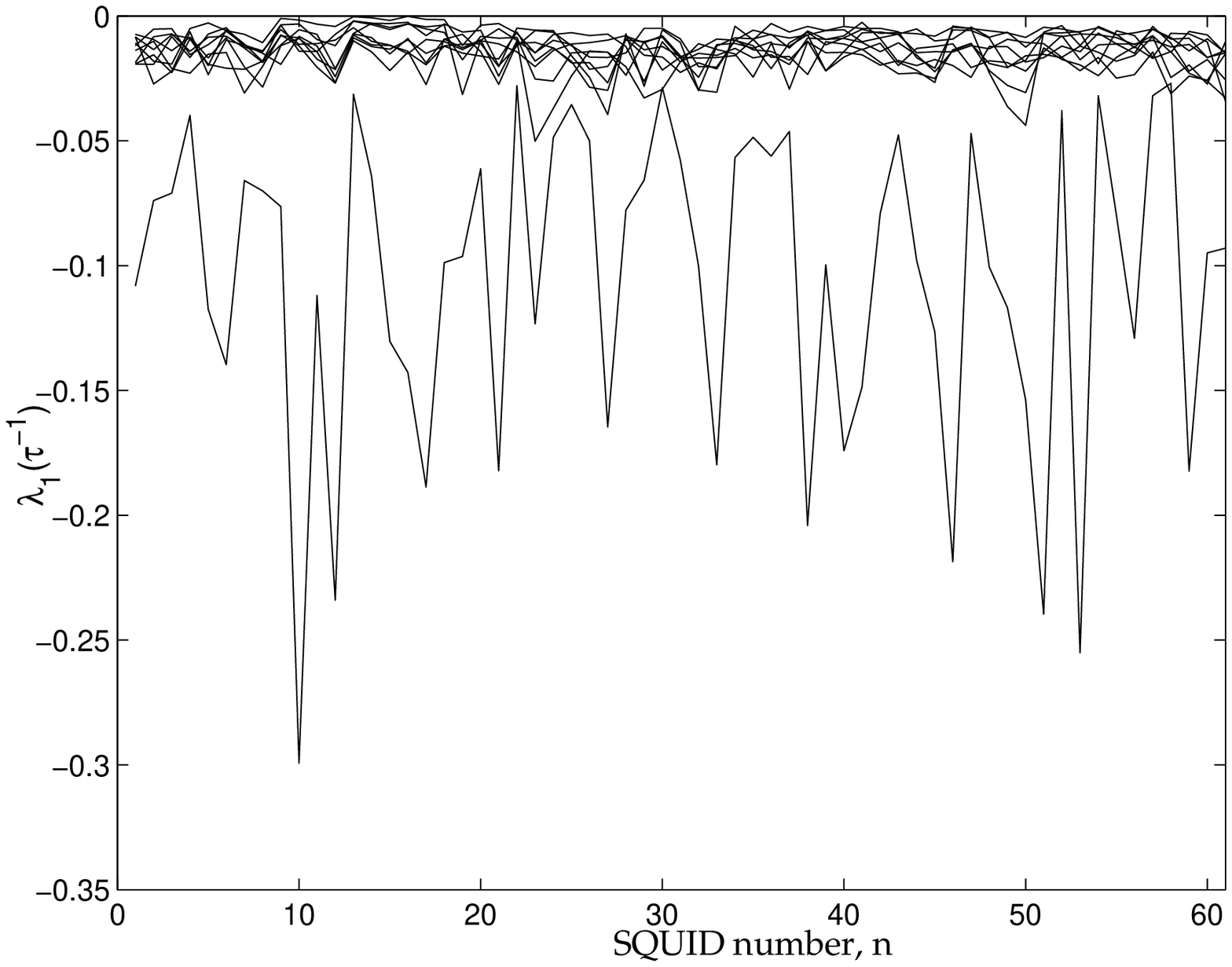}
\caption{The  topographic dependence of the first relaxation
parameter $\lambda_1$ for nine healthy subjects (upper lines) in
comparison with the  patient with PSE (lower line) for an R/ B
combination of the light stimulus. The crucial role of the brain
zones with sensor numbers m = 10, 12, 46 , 51 and 53 is clearly
visible.}
\end{figure}


 Thus, the signals in the patient with PSE with   sensor number 10
  differ from the healthy subjects consists in the
drastic change of the fluctuation scales of the dynamic orthogonal
variables and the space sizes of the phase clouds. The similar
difference of the scales constitutes the values from 10 to 80 times.
This observation let us note the specific role and behavior of the
sensor with  number n=10 in the formation of PSE mechanisms! The
difference in the scales of the orthogonal dynamic variables and in
the sizes of the phase clouds  is more drastic for sensors with
numbers n = 10, 5, 23, 40 and 53.

\begin{figure}[h!]
\leavevmode \centering
\includegraphics[height=9cm,width=14cm,angle=0]{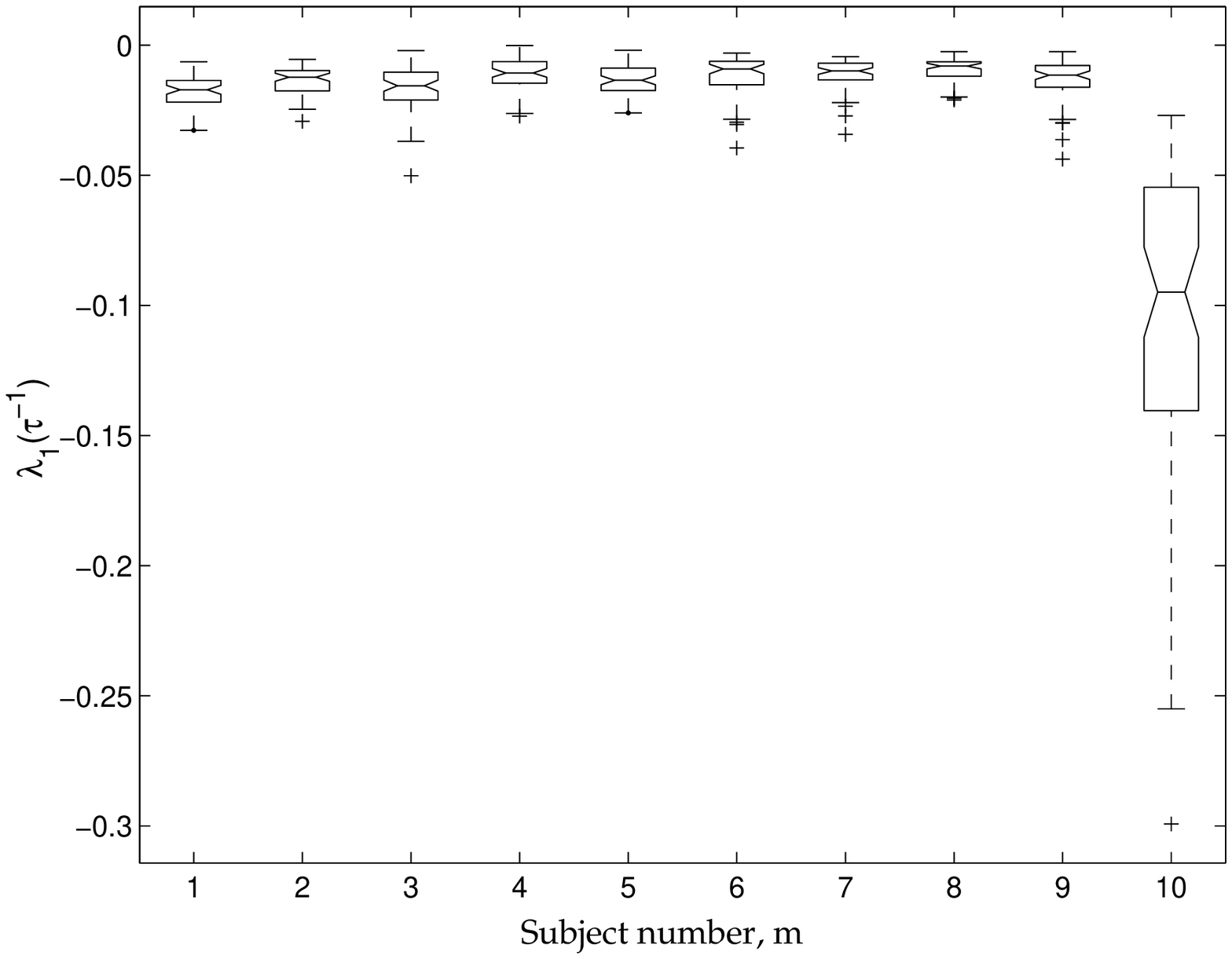}
\caption{The  mean values  of the first relaxation parameter
$\lambda_1$ for the whole  group of the  nine healthy subjects (n =
1, 2, 3  ... 9), averaged on the total set of sensors m = 1, 2,
3,... 61 versus the patient with PSE (m = 10) for an R/ B
combination of the light stimulus. One can note the drastic
difference of approximately of  4,4 times for the  healthy vs the
patient with PSE!}
\end{figure}



 From the time dependence of the initial TCF $ M_0(t)$ and the first three MF's of the junior
order $M_i(t),$ i = 1, 2 and 3  it is possible to see great
difference in the behavior of the time functions for the healthy and
for the patient with PSE.  One can observe a large-scale time
correlation in the healthy subjects in the time dependence of MF's
$M_i(t), i = 0, 1, 2$ and 3, whereas the similar functions
demonstrate  a small-scale fluctuation and a small-amplitude
oscillation in  case of the patient with PSE.

\begin{figure}[h!]
\leavevmode \centering
\includegraphics[height=9cm, width=14cm, angle=0]{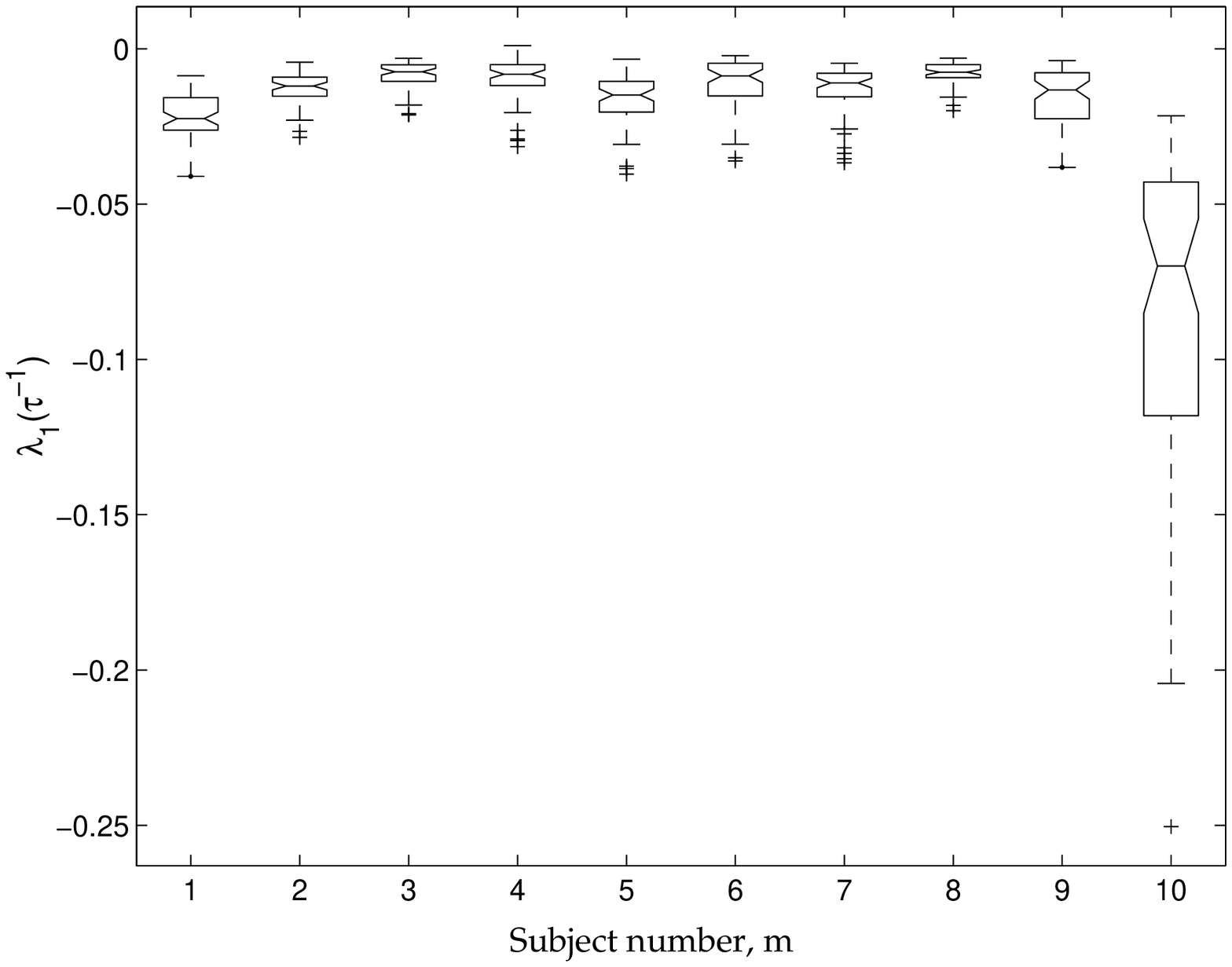}
\caption{The mean values of the first relaxation parameter
$\lambda_1 $ for the group of the nine healthy subjects (m=1, 2,
...9) averaged on the whole set of sensors  with numbers $ 1 \leq n
\leq 61$ vs the patient with PSE (m= 10) for an R/ G combination of
the light stimulus. The distinction between the healthy and the
patient with PSE amounts up to 8 times!}
\end{figure}

One can note that the sensors with  numbers n = 10, 5, 23, 40 and
53 are specific points in the brain core of the patient with PSE.
It is interesting to  observe a dynamical picture for the usual
and the nonspecific points at the human cerebral cortex. To this
aim, the results for the nonspecific  sensor with number n = 13
are submitted  in Figs. (2) - (6). Figs. (2) (for the healthy) and
Figs. (5) (for the patient with PSE) present the time dependence
of the first four dynamical orthogonal variables. One can detect a
smooth  behavior of the variables for $\ W_i(t) $ in the  healthy
and, in clear contrast, a sharp or irregular dynamics of $\ W_i(t)
$ in the patient with PSE. Figs. (3) (for the  healthy) and Figs.
(6) (for the patient with PSE) show the construction of the phase
space made up by separate phase points. We depict stratified phase
space for the healthy person and for the patient with PSE. Figs.
(4) show the nonlinear dynamics of the formation of the phase
space by the phase trajectories for the healthy.

Here one can observe the pseudoperiodic orbital movement for the
phase trajectory in the healthy and the quasi-strange attractors in
the patient with PSE. In this context one must note  the small time
scales in the dynamics in the  healthy and the larger  time scales
in nonlinear dynamics in the patient with PSE. Figs. (7)  depict the
time dependence of the initial TCF and the first three memory
functions $M_i(t),$ i= 1,2 and 3 for the patient with PSE. Large
scale fluctuations and oscillations are visible in the time
dependence of $M_i(t),$ i = 0, 1, 2 and 3 in the healthy and small
scale deformation are evident in the patient with the PSE.

    Figs. 8 show the topographic dependence as a function of SQUID-number of the first
relaxation parameters $\lambda _1$  for a red-blue (R/B)
combinations of the light stimulus for the healthy subjects in
comparison with the patient with PSE. Similar results one can see
for a red-green (R/G) combinations of the light stimulus.
 One can observe a large difference of the numerical values of this
parameter in the  healthy as compared to the patient with PSE.  The
parameter $\lambda _1$ differs on average 6-7 times in the all the
sensors. One can note a specially strong difference in the data
between the healthy and the patient in numerical values of parameter
$\lambda _1$ especially in the sensors with  numbers n = 10, 12, 46,
51 and 56 (for an R/ B combination of the light stimulus) and n =
10, 12, 46, 51 and 53 (for an R/ G combination of light stimulus).
Both combination of the light stimulus (an R/B and an R/G) yield
approximately identical results for the sensor's location.

   Figs. 9 (for an R/ B combination of the light stimulus) and 10
   (for an R/ G combination) demonstrate the behavior of relaxation parameter
$\lambda _1$ for each individual  of nine healthy subjects averaged
on  all sensor location in cerebral cortex in comparison with the
patient with PSE. A  remarkable difference of $\lambda _1$ appears
in healthy subjects, this being approximately 4 - 8 times, on
average 7 times for an R/ B combination in Fig. 9,  and 4, 4 times
for an R/G combination of the light stimulus in Fig. 10. This
difference is a reliable indicator of serious destruction in
functioning of the human organism with PSE. It is important that the
behavior of the coefficients of $\lambda _1$ specifies the
singularities of relaxation mechanisms in the MEG's signals. From
the physical point of view parameter $R_1 =|\lambda _1|$ mimics a
relaxation rate. We see the drastic distinction in relaxation rates
for  a healthy person as compared to the patient with PSE. This fact
indicates the crucial role of the specific relaxation processes in
the pathological functioning of the human cerebral cortex for the
patient suffering from PSE.

  One  notes that the sensors with the  numbers  n= 10,
  12, 46 and 51 locate specific points  in the patient's brain  with PSE.
   It is necessary to
mention the overall positions of the sensors. This indicates that
the potential abnormality is not confined to the occipital region
but distributed over the entire brain.

\section{Conclusions}

Our discussion in this work make it evident that physiological
models typically possess a great number of parameters, each with
their natural range of variability and uncertainty in measurement.
Relaxation and dynamic behavior of the system signals can vary
widely from one chosen set of parameters to another. The presented
information-theoretic memory-function analysis provides one
possibility of extracting interrelations within this complicated
parameter space. The study of the physical and dynamical boundaries
between different types of behavior is a necessity, both for a
better understanding of brain function and for the application of a
diagnosing procedure and treatment of patients suffering PSE.

Control can be applied at preventing the brain from entering an
undesirable, pathological state such as a seizure \cite{Parra}.
Here we have shown that the parameter space of MEG activity in the
patient with PSE gives rise to a robust chaotic behavior.
 In order to study spatiotemporal
cortical dynamics we need to analyze the global MEG data. In this
paper we could cognize that the relaxation and dynamic singularities
may account for the registration of the relevant pathological zones
in the human cerebral cortex which are responsible for epilepsy.

Many natural phenomena can be described by distributions exhibiting
a time scale-invariant behavior \cite{Stanley,Worrell}. The
suggested approach allows the stochastic dynamics of neuromagnetic
signals in human cortex to be treated in a statistical manner and to
search for its characteristic statistical identifiers. From the
physical point of view, the obtained results  can be put to use as a
possible test to identify the presence or absence of brain anomalies
as they occur in a patient with PSE. The set of our quantifiers is
uniquely associated with the emergence of time-scale and relaxation
effects in the chaotic behavior of the neuromagnetic responses of
the human brain core. The registration of the behavior of those
indicators discussed here is thus of beneficial use in order to
detect pathological states of separate areas (sensors 10, 5,23, 40
and 53 in our case under consideration) in the human brain of a
patient with PSE.

PSE is a type of reflexive epilepsy which originates mostly in
visual cortex (both striate and extra-striate) but with high
possibility towards propagating to other cortical regions
\cite{Binnie}. Healthy brain may possess an inherent controlling (or
defensive) mechanism against this propagation of cortical
excitations, the breakdown of which makes the brain vulnerable to
trigger epileptic seizures in patients \cite{Porciatti}. However,
the exact origin and dynamical nature of this putative defensive
mechanism is not fully known. Earlier we showed \cite{MEG} that
brain responses against chromatic flickering in healthy subjects
represent strong nonlinear structures whereas nonlinearity is
dramatically reduced to minimal in patients.

 There are
other quantifiers of a different nature, such as the Lyapunov's
exponent, Kolmogorov-Sinai entropy, correlation dimension, etc.,
which are widely used in nonlinear dynamics  and related
applications, see in Ref. \cite{kantzschreiber}. In the present
context, we find that the employed statistical and dynamical
measures are not only convenient for analysis, but are also
suitable for identification of anomalous brain behavior. The
search for other quantifiers, and foremost, the optimization of
such measures when applied to complex discrete time dynamics
presents a true challenge. This objective particularly holds true
when attempts are made to identify and quantify  an anomalous
functioning in living systems. This study presents a first
stepping stone towards understanding of nonlinear brain processes
defending against hyper excitation to flickering stimulus by new
analysis techniques based on non-Markov random processes.




\acknowledgments
 This work was supported by the Grants of RFBR  $N$
05-02-16639a, Ministry of Education and Science of Russian
Federation $N$ 2.1.1.741 (R. Y., D. Y., and E. K.) and JST.
Shimojo ERATO project (J. B. , K. W. , and S. S.).

\end{document}